\documentclass[12pt]{article}
\usepackage[english]{babel}

\usepackage{amsmath}
\usepackage{amssymb}
\usepackage{graphicx,bm}
\usepackage{spverbatim}
\usepackage{authblk}
\usepackage{subcaption}
\usepackage{caption}
\usepackage{float}
\usepackage{cite}
\DeclareCaptionLabelSeparator{periodspace}{.\quad}
\title{Cross-Correlation in cricket data and RMT}
\date{\today}
\author {Manu Kalia \footnote{Corresponding author: manukalia24@gmail.com} }

\author{Saugata Ghosh}
\affil{The Creative School, E-791, C. R. Park, New Delhi-110019, India}

\begin{document}
    \maketitle
    \begin{abstract}
    We analyze cross-correlation between runs scored over a time interval in cricket matches of different teams using methods of random matrix theory (RMT). We obtain an ensemble of cross-correlation matrices $C$
from runs scored by eight cricket playing nations for (i) test cricket from 1877 -2014 
(ii)one-day internationals from
1971 -2014 and (iii) seven teams participating in the Indian Premier league T20 format (2008-2014) respectively.
We find that a majority of the eigenvalues of C fall within the bounds of random matrices having joint probability distribution $P(x_1\ldots,x_n)=C_{N \beta} \, \prod_{j<k}w(x_j)\left | x_j-x_k \right |^\beta$ where $w(x)=x^{N\beta a}\exp\left(-N\beta b x\right)$ and $\beta$ is the Dyson parameter. The corresponding level density  gives Marchenko-Pastur (MP) distribution while fluctuations of every participating team agrees with the universal behavior of Gaussian Unitary Ensemble (GUE). We analyze the components of the deviating eigenvalues and find that the largest eigenvalue corresponds
to an influence common to all matches played during these periods.
    \end{abstract}
%%%%%%%%%%%%%%%%%%%%%%%%%%%%%%%%%%%%%%%%%%%%

PACS numbers: 05.45.Tp, 05.40.-a

\section{Introduction}
Analyzing correlations among cricket teams of different era has been a topic of interest for sports experts and journalists for decades. In this paper we study such influence (or interaction) by constructing cross-correlation matrix $C$ \cite{plerou,fin,fin1,fin2,fin3,quantrans} formed by runs scored by teams over different time intervals, formally called a time series.We consider the time series of batting scores posted per innings by a team in all official ICC International Test matches played. Then we construct an ensemble of cross-correlation matrices corresponding to Test data for that cricket team. We repeat the process for One Day International (ODI) and Indian Premier League (IPL) T20 cricket matches. We assume
the correlations to be random and compare the fluctuating properties of $C$ with that of random matrices. Within the bounds imposed by the RMT model, fluctuations of
$C$ show brilliant agreement with the ``universal'' results of GUE \cite{mehta,ghosh,ghoshbook}, while the level density corresponds to the MP distribution \cite{marchenko}. This implies that
interactions in $C$ are random, or in simple words not governed by any causality principal. However outside the bounds, eigenvalues of $C$ show departure from RMT predictions, implying
influence of external non-random factors common to all matches played during this period. To understand this effect, we remove $k$ extreme bands from $C$ and perform the Kolmogorov-Smirnov (KS) Test. We observe a better agreement with RMT predictions.

%%****************************************

We organize the paper as follows: After a brief description of the data analyzed in sub-section [\ref{sec:data}], we define cross-correlation matrix in sub-section[\ref{sec:acm}]. Section[\ref{sec:rmt}] introduces our RMT model along with a brief proof of MP distribution. We analyze our results and its corresponding RMT model in Section [\ref{sec:analysis}]. This is followed by concluding remarks.

%%%%%%%%%%%%%%%%%%%%%%%%%%%%%%%%%%%%%%%%%%%%%%%%%%%%%%%%%%%%%%%%%%%%%%%%%%%%%%

\subsection{Data analysed}
\label{sec:data}
We construct three ensembles, corresponding to runs scored in Tests, ODIs and Indian Premier League (IPL).

\begin{itemize}
\item The ODI ensemble comprises of cross-correlation matrices constructed from runs scored by India, England, Australia, West Indies, South Africa, New Zealand, Pakistan and Sri Lanka for all official ICC One Day International matches played between 1971 and 2014. For each country we have a sequence of runs scored in both home and  away matches. An ensemble of fifty one  $90\times 90$ matrices are constructed from the time series data.

\item The Test ensemble comprises of cross-correlation matrices constructed from runs scored by India, England, Australia, West Indies, South Africa, New Zealand, Pakistan and Sri Lanka. For each country we have a sequence of runs scored per innings (each match has a maximum of two innings) in  both home and away matches. The Test scores have been taken for all matches played between England, Australia and South Africa between 1877 and 1909 and all official ICC Test matches thereafter, till 2014. An ensemble of seventy $90 \times 90$ matrices are constructed from the time series data.

\item The IPL ensemble comprises of cross-correlation matrices constructed from runs scored by Chennai Super Kings, Rajasthan Royals, Royal Challengers Bangalore, Delhi Daredevils, Kings XI Punjab, Kolkata Knight Riders and Mumbai Indians for all official BCCI IPL T20 matches played between 2008 and 2014. For each team we have a sequence of batting scores posted per match. An ensemble of twenty eight $20 \times 20$ matrices are constructed from the time series data.
\end{itemize}

%%%%%%%%%%%%%%%%%%%%%%%%%%%%%%%%%%%%%%%%%%%%%%%%%%%%%%%%%%%%%%%%%%%%%%%%%%%%%%%%%%%%%%%%%%%%%%

\subsection{Cross-correlation matrix}
\label{sec:acm}
Cross-correlation matrix $C$ is constructed from a given time series $X=\left\lbrace X(1),X(2),\ldots\right\rbrace$ by defining subsequences
$X_{i}=\left\lbrace X(i),X(i+1),\ldots,X(N) \right\rbrace$ and
$X_{j}=\left\lbrace X(j),X(j+1),\ldots,X(N-\Delta t) \right\rbrace$,
separated by a ``lag'' $\Delta t = i-j$, $j<i$ and $i,j \in \mathbb{N}$.
We then normalize the subsequences by defining
\begin{equation}
Y_i=\frac{X_i - \mu_{X_i}}{\sigma _{X_i}}.
\end{equation}
Finally, cross-correlation matrix $C$\cite{plerou} is defined as
\begin{equation}
C_{i,j}=\left< Y_i Y_j \right>,
\end{equation}
where $\mu_{X_i}$ and $\sigma_{X_{i}}$ are sample mean score and standard deviation of the subsequence $X_i$ respectively, and $\left<\ldots\right>$ denotes a time average over the period studied. This is the correlation coefficient between the subsequences $Y_i$ and $Y_j$ and help us understand the correlation between runs scored by a given team at different time intervals. The matrix elements lie between -1 and 1 and the matrices so constructed are Hermitian. 

Now, we construct multiple matrices on a single time series, giving rise to an ensemble of matrices. Letting $C^{(1)}=C$ (as constructed above), we construct another matrix $C^{(2)}$ by removing first $N$ elements of the time series considered, and constructing the cross-correlation matrix with the method described above. We continue this process of construction till the length of the truncated time series becomes less than $N$.

%%%%%%%%%%%%%%%%%%%%%%%%%%%%%%%%%%%%%%%%%%%%%%%%%%%%%%%%%%%%%%%%%%%%%%%%%%%%%%

\section{Random Matrix Model}
\label{sec:rmt}

Unitary Ensemble of random matrices is invariant under unitary transformation $H\rightarrow W^{T}HW$ where the ensemble is defined in the space $T_{2G}$ of Hermitian matrices and $W$ is any unitary matrix. Also, the various linearly independent elements of $H$, must be statistically independent\cite{mehta}.

Joint probability distribution function of eigenvalues $\{ x_1,x_2,...,x_N \}$  is given by,
\begin{equation}
\label{jpdf}
P_{N\beta}(x_1,..,x_N)=C_{N \beta}. \, \prod_{j<k}x_{j}^{N\beta a}\exp\left(-N\beta b \sum_{1}^{N}x_j\right)\left | x_j-x_k \right |^\beta,
\end{equation}
where $\beta=1,2$ and $4$ correspond to orthogonal (OE), unitary (UE) and symplectic (SE) ensembles respectively and $C_{N\beta}$ is the normalization constant \cite{mehta}.
We define $n$-point correlation function by
\begin{equation}
R_{n}^{(\beta)}(x_1,..,x_n)=\frac{N!}{(N-n)!}\int dx_{n+1}\ldots\int dx_{N}P_{N\beta}(x_1,..,x_N).
\end{equation}
This gives a hierarchy of equations \cite{ghoshbook} given by
\begin{eqnarray}
\label{hier}
\beta R_{1}(x)\int\frac{R_{1}(y)}{(x-y)}dy+\frac{w\prime(x)}{w(x)}R_{1}(x)=0,
\end{eqnarray}
where
\begin{equation}
w(x)=x^{N\beta a}\exp[{-N\beta b x}].
\end{equation}

\begin{figure}[H]
        \centering
        \includegraphics[width=0.7\textwidth]{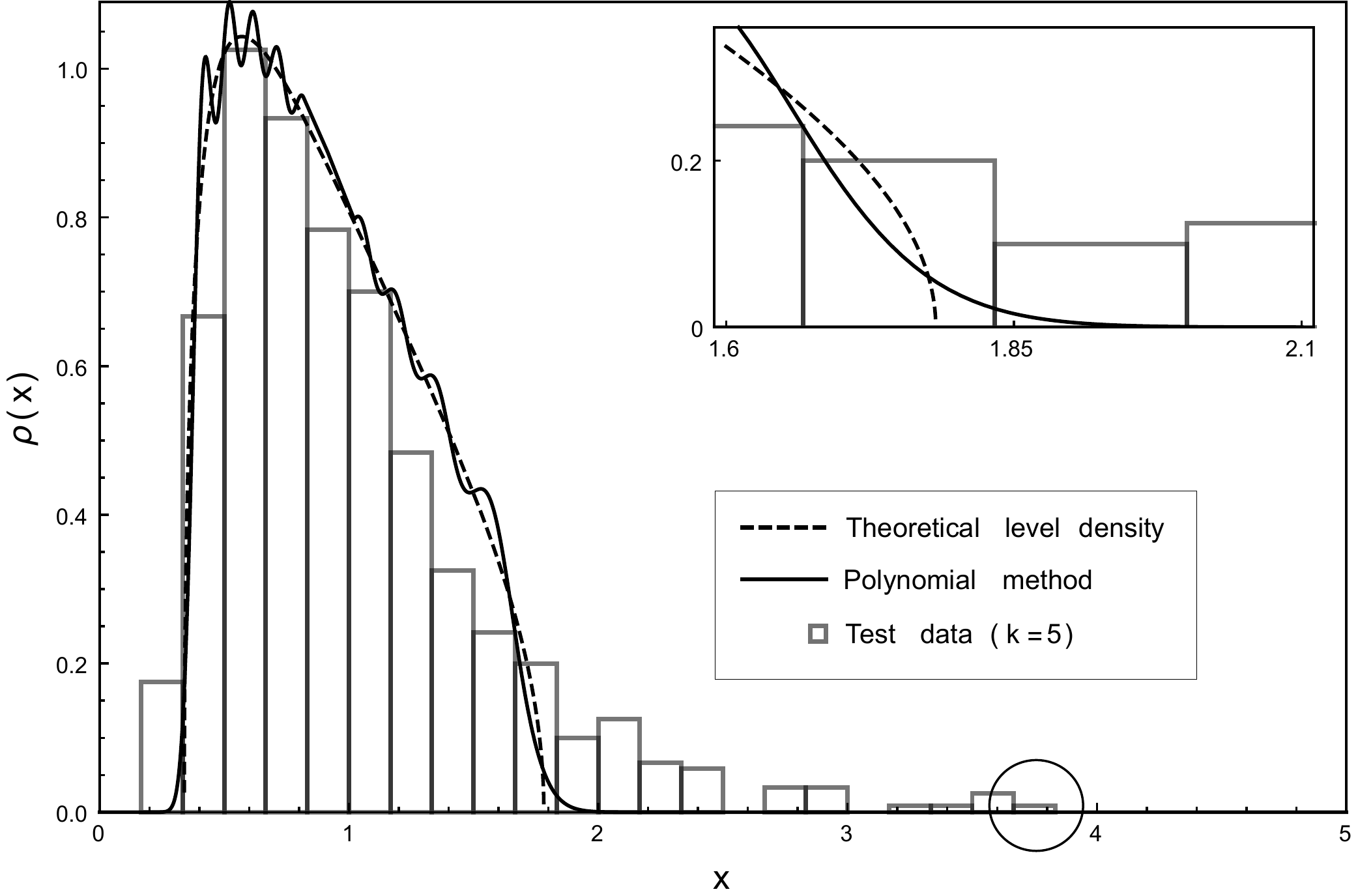}
         \caption{Level Density for averaged Test data with $k=5$ . The solid line refers to Marchenko-Pastur result  (\ref{den(th)}) and the dashed line refers to the finite $N$ result, obtained by the polynomial method described in Section \ref{sec:rmt}. Here, $a=2.75$, $b=3.535$, $X_{-}=0.339601$ and $X_{+}=1.78204$ in (\ref{den(th)}). The largest eigenvalue is circled towards the end of the spectrum.}\label{ld}
\end{figure}

We solve the integral equation using the resolvent
\begin{equation}
G(z)=\int\frac{R_{1}(y)}{z-y}dy,
\end{equation}
which satisfies
\begin{equation}
G(x+i0)=\int\frac{R_{1}(y)}{x-y}dy-i\pi R_{1}(x).
\end{equation}
Multiplying Eq.(\ref{hier}) by $x/(z-x)$ and integrating over $x$ we get after some elementary calculation
\begin{eqnarray}
\label{den(th)}
\rho(x)\equiv \frac{R_{1}(x)}{N} &=& \frac{b}{\pi x}\sqrt{(x-X_{-})(X_{+}-x)};\hspace{1cm} X_{-}<x<X_{+},\\
 \nonumber
                                 &=& 0, \hspace{2cm}\textrm{otherwise.}
\end{eqnarray}
where
\begin{equation}
X_{\pm}=\frac{a+1}{b}\pm \frac{\sqrt{2a+1}}{b}.
\end{equation}

For finite $N$, following Dyson-Mehta method \cite{mehta}, we  use
\begin{equation}
\label{den(fin)}
\rho(x)=\frac{1}{N}\sum_{j=0}^{N-1}\phi^{2}_{j}(x),\hspace{1cm}\phi_{j}(x)=\sqrt{w(x)}P_{j}(x),
\end{equation}
where $P_{j}(x)$ are orthonormal polynomials which satisfy
\begin{equation}
\int_{X_{-}}^{X_{+}}P_{j}(x)P_{k}(x)w(x)dx=\delta_{j,k},\hspace{1cm}j,k\in \mathbb N.
\end{equation}

To understand the correlation in the system, we first need to unfold the eigenvalues to eliminate global effect over fluctuation. The sequence of scores for each country is unfolded independently. The corresponding unfolded eigenvalues $y_k$ are given by\cite{verbaarschot},
\begin{equation}\label{unf}
y_k=\int_{X_-}^{x_k}\rho(x)\mathrm{d}x,
\end{equation}
and the mean spacing of the unfolded eigenvalues $y_k$ is 1. We perform unfolding using both (i) the theoretical level density (\ref{den(th)}) and (ii) numerical integration of the data and obtain the best-fit over the integrated density.

\begin{figure}[H]
    \centering
        \begin{subfigure}[b]{0.47\textwidth}
        \includegraphics[width=\textwidth]{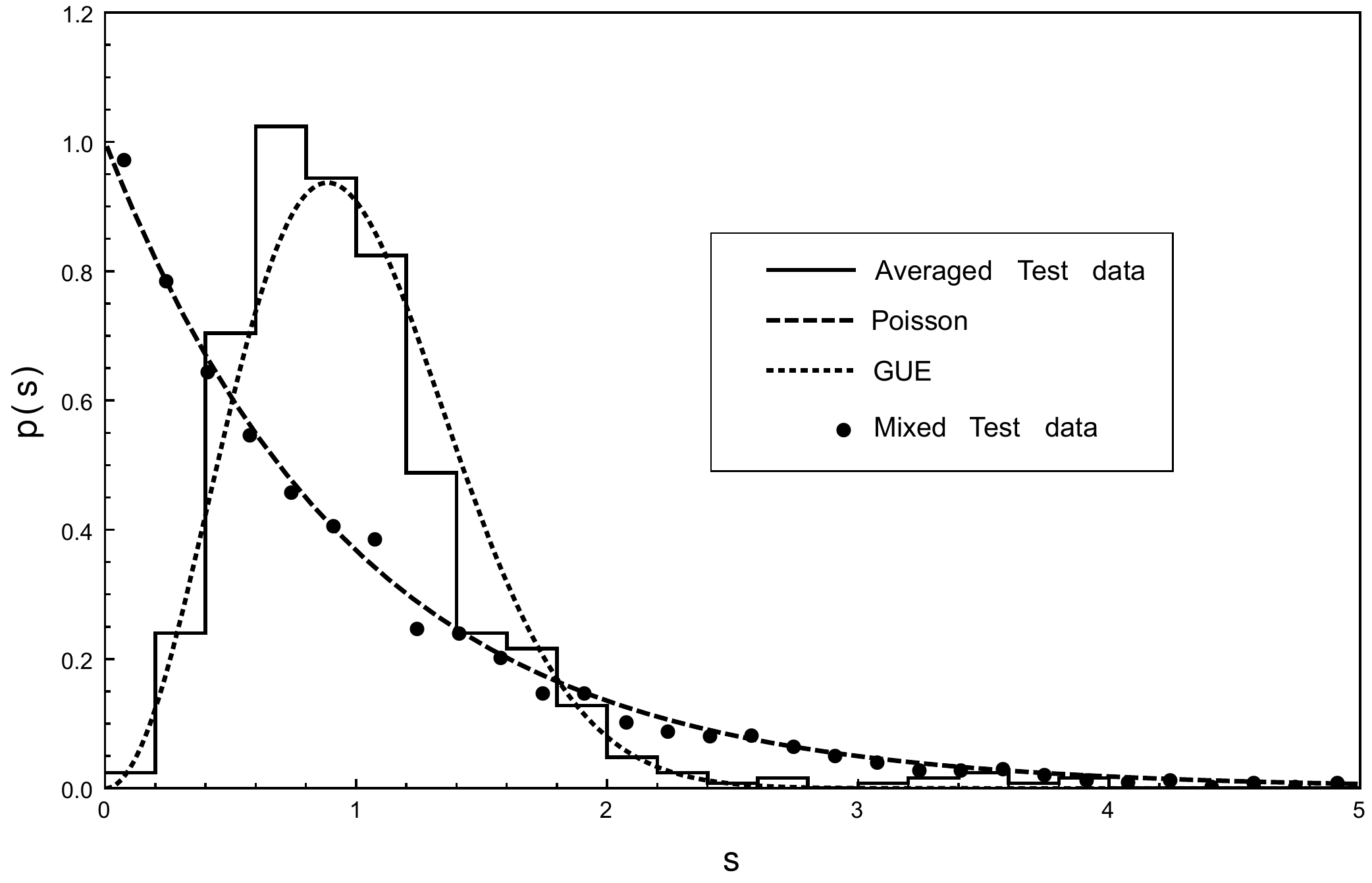}
        \caption{Theoretical unfolding}
    \end{subfigure}\hfill
    \begin{subfigure}[b]{0.47\textwidth}
        \includegraphics[width=\textwidth]{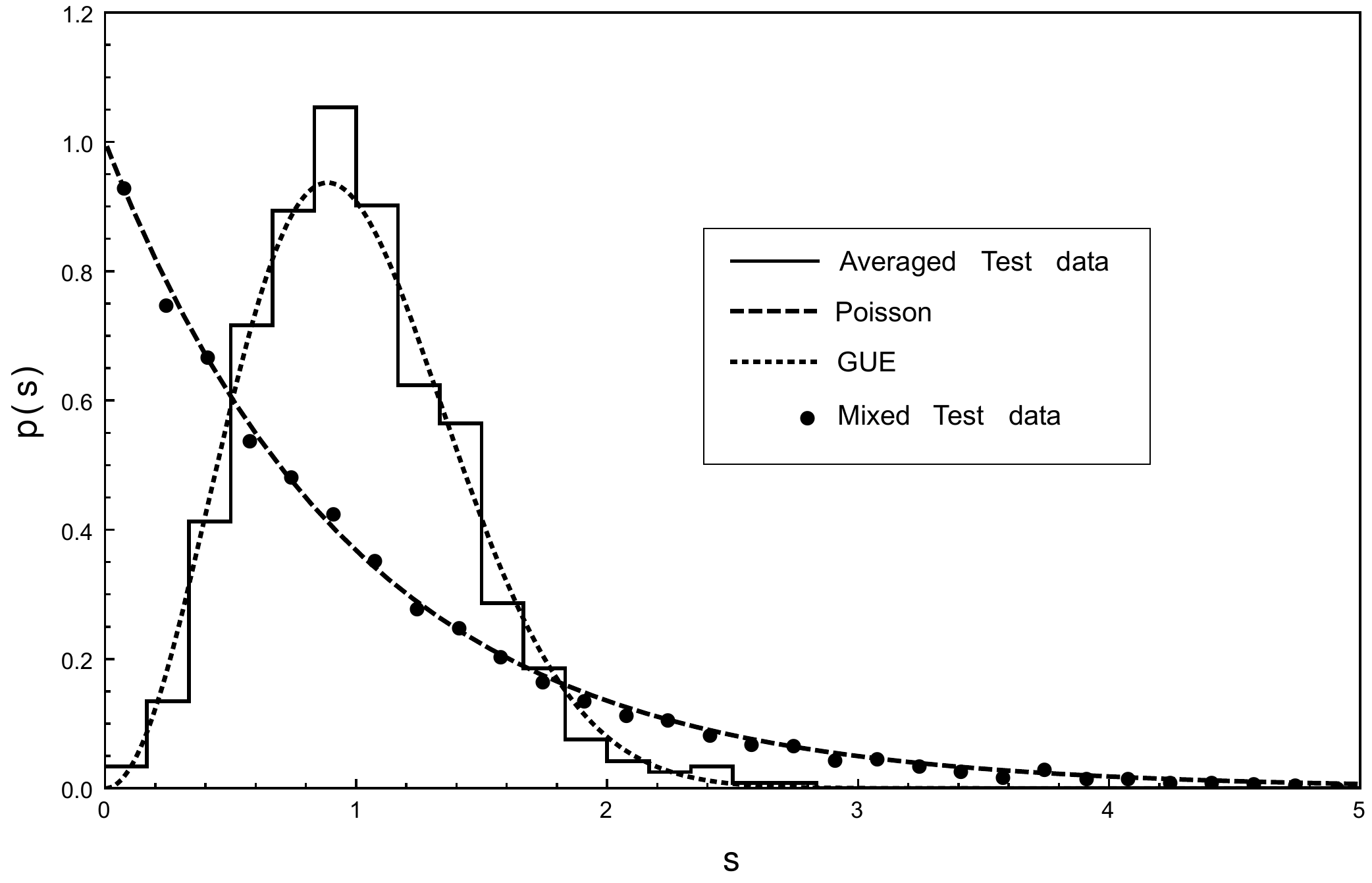}
        \caption{Numerical unfolding}
    \end{subfigure}
    \caption{Nearest neighbour spacing distribution for mixed and averaged Test data obtained via numerical and theoretical unfolding (using Marchenko-Pastur result (\ref{den(th)}) with $a=2.75$, $b=3.535$, $X_{-}=0.339601$ and $X_{+}=1.78204$). The solid line refers to spacing distribution of experimental data with $k=5$, the dotted line refers to GUE result and the dashed line refers to the Poisson case.}\label{sd}
\end{figure}

For $\left \{S_i | S_i=y_{i+1}-y_i  \right \}$, $s_i=S_i/D$ where $y_i$ denote successive unfolded levels and $D$ is the average spacing, the level spacing distribution $p(s)ds$ is defined as the probability of finding an $s_i$ between $s$ and $s+ds$ \cite{mehta}.
 For no correlations between the levels, we have the Poisson distribution
    \begin{equation} \label{poisson}
  p(s)=\exp[-s],
  \end{equation}
while for GUE, we get the Wigner's surmise
 \begin{equation} \label{guespacing}
  p(s)=\frac{32 s^2}{\pi^2} \exp \left[-\frac{4}{\pi} s^2\right].
\end{equation}

We consider $8$ sequences of eigenvalues for Test data obtained by ensemble averaging over each country. We  unfold these sequences individually and average over the $8$ sequences of spacings. The result shows remarkable agreement with GUE predictions (Fig. \ref{sd}). Upon mixing of the eigenvalues of the Test data we observe Poisson distribution (Fig. \ref{sd}).

\begin{figure}[H]
        \centering
        \includegraphics[width=0.7\textwidth]{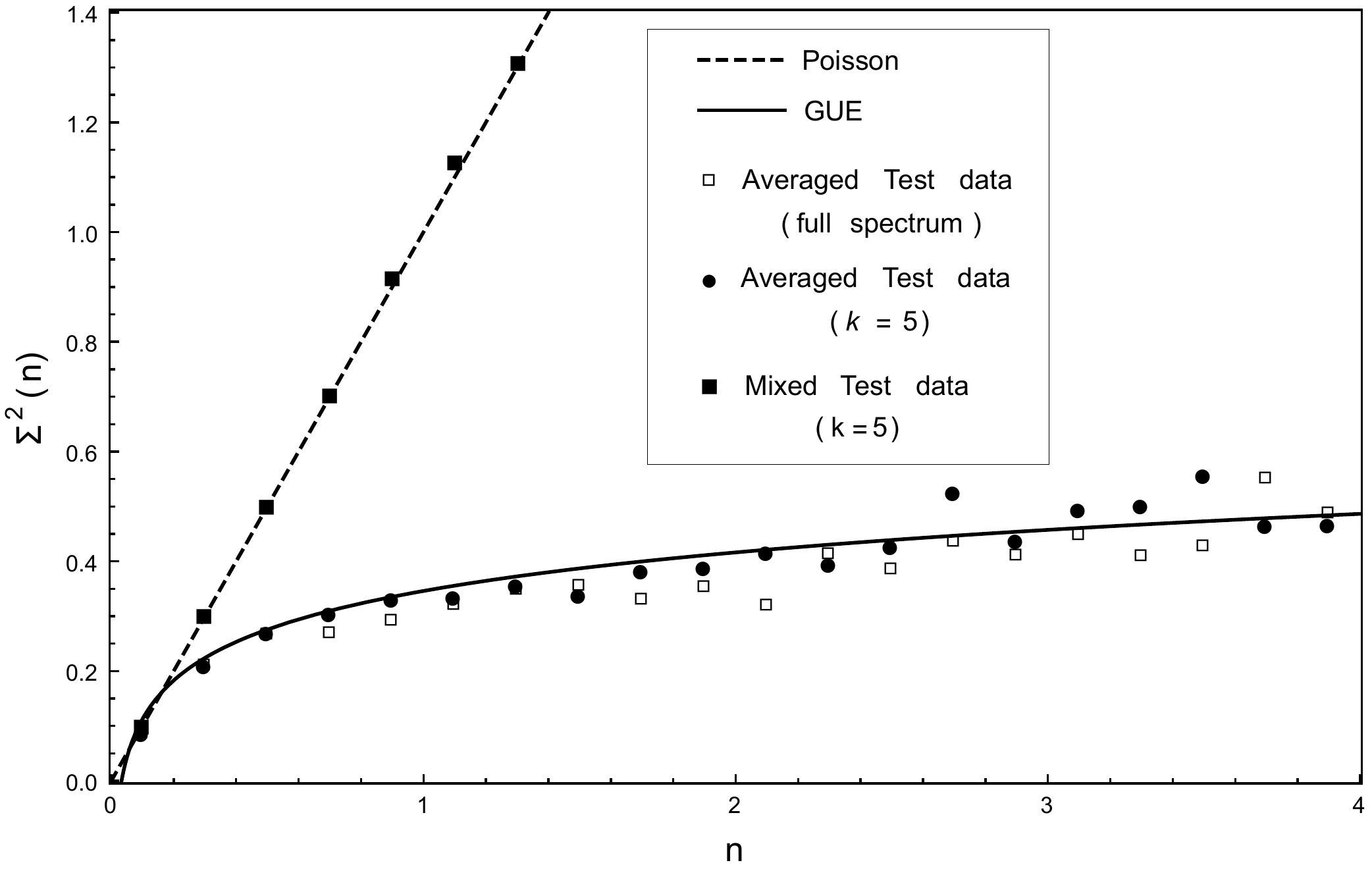}
         \caption{Number variance for the averaged and mixed Test data obtained via numerically unfolding over the spectra. The solid line refers to GUE result (\ref{numbervariance}) and the dashed line refers to Poisson case. The figure plots three cases: (i) Averaged Test data with $k=5$ extreme diagonals removed (ii) Mixed Test data with $k=5$ extreme diagonals removed and (iii) Mixed Test data for the entire spectrum when no diagonals are removed from the matrices.}\label{nv}
\end{figure}

   Another statistic considered is the linear statistic or the number variance. For $n_k$ unfolded levels in consecutive sequences of length $n$, we define the moments \cite{verbaarschot},
\begin{equation}   
M_p(n)=\frac{1}{N}\sum_{k=1}^{N}n_k^p,
\end{equation}
    where $N$ is the number of sequences considered, each of length $n$ covering the entire spectrum. Then the number variance $\Sigma^2(n)$ is given by
\begin{equation}
\Sigma^2(n)=M_2(n)-n^2.
\end{equation}
    For GUE, number variance is given by \cite{mehta},
\begin{equation} \label{numbervariance}
    \Sigma^2(n)=\frac{1}{\pi^2}\left(\ln(2\pi n)+ \gamma +1 \right),
\end{equation}
where $\gamma$ is the well known Euler constant. Number variance is known to be very sensitive for larger values of $n$  on account of spectral rigidity. Fig \ref{nv} shows a very good agreement of the experimental number variance result of the Test data to that of the GUE result for cases when $k=0$ and $k=5$ extreme diagonals are removed from both ends of the matrices involved in calculation.

\begin{figure}[H]
        \centering
        \includegraphics[width=0.7\textwidth]{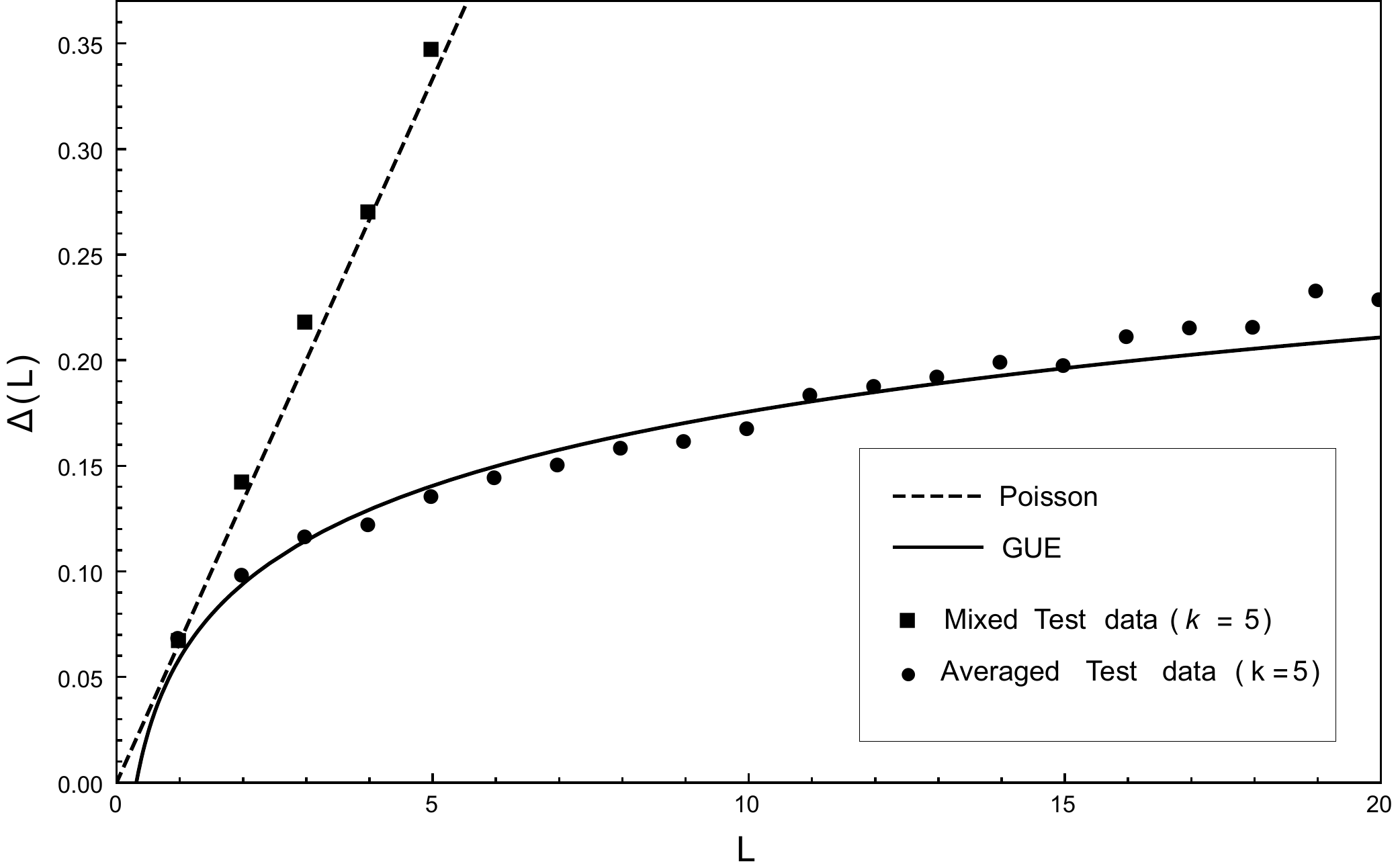}
         \caption{The Dyson-Mehta least squares statistic for the averaged and mixed Test data with $k=5$ extreme diagonals removed from both ends of the matrices involved in calculation obtained via numerically unfolding the spectrum . The solid line refers to the GUE result (\ref{guedel}) and the dashed line refers to the result for the Poisson case (\ref{poisdel}).}\label{del}
\end{figure}

The other statistics considered is the Dyson-Mehta least square statistic or the spectral rigidity statistic\cite{mehta} which measures the long-range correlations and irregularity in the level series in the system by calculating the least square deviation of the unfolding function from a straight line $y=a E + b$ over different ranges $L$. The statistic $\Delta(L)$ for $L=L_2-L_1$ is given by the integral,
\begin{equation}\label{delta}
\Delta (L)=\frac{1}{L}\int_{L_1}^{L_2}(N(E)-aE-b)^2dE,
\end{equation}
where $N(E)$ is the unfolding function. The mean value of the statistic for the GUE case is given by \cite{mehta},
\begin{equation}\label{guedel}
\left \langle \Delta \right \rangle = \frac{1}{2 \pi^2} (\ln (2 \pi L) + \gamma - 5/4).
\end{equation}
For Poisson case, the least square statistics is given by
\begin{equation}\label{poisdel}
\left \langle \Delta  \right \rangle = \frac{s}{15}.
\end{equation}

%%%%%%%%%%%%%%%%%%%%%%%%%%%%%%%%%%%%%%%%%%%%%%%%%%%%%%%%%%%%%%%%%%%%%%%%%%%%%%%%%%%%%%%%%%%%%%%%%%%%%%%%%

\section{Analysis}
\label{sec:analysis}

The problem that one encounters in analysis of such data are 

1. The finite length of time series available introduces measurement noise.

2. A bigger time series will introduce more contributions from non-random events which will affect the ``universality'' result
    but will provide information about the correlations among different time series.
    
We study the RMT model defined by Eq.(\ref{jpdf}). We obtain MP distribution (\ref{den(th)}) for the level density as $N\rightarrow \infty$. We observe that
the level density of eigenvalues of $C$ in the bulk shows a remarkable agreement with the MP distribution for all Test, ODI and IPL data. However, some large eigenvalues exist outside the bounds $[X_-, X_+]$. To ensure that these eigenvalues are not due to finite $N$ effect, we obtain level-density for finite $N$. 
For this, we develop the corresponding orthonormal polynomials using Gram-Schmidt method and using  Eq.(\ref{den(fin)}) for $N=10$ obtain the level density and
 compare that with ensembles of cricketing data. (Fig. \ref{ld}). We observe that the large eigenvalues still remain outside the bounds.\\

The next question is if these large eigenvalues non random, in which case our RMT model will not only show disagreement with the level density but also ``spoil'' the RMT predictions. To verify this, we make RMT analysis over the entire spectrum and compare its results with the truncated sparse matrix, which removes the large eigenvalues. KS test shows that our level density and spacing distribution analysis is considerably hampered by the presence of these large eigenvalues, thereby conforming the existence of non random long range correlations.  

 To track the level of non-randomness, we remove $k$, ($k<<N$) extreme bands out of $2N-1$ bands of the $N\times N$ matrices $C$ and perform the KS test. We perform numerical unfolding over the eigenvalues where the integrated density of states are fitted with a polynomial. For ODI, where $N=90$ we obtain a p-value of $0.640311$ for the full spectrum and a p-value of $0.9025$ for spectrum of the matrix with $k=15$. For the Test data (again $N=90$), we obtain a p-value of $0.49$ for unfolding the full spectrum and a p-value of $0.855394$ when unfolding the spectrum of the matrix with $k=5$.Thus
by creating a sparse matrix, which removes the large eigenvalues, our results converge to RMT predictions by $\approx30 \%$. This proves the existence of non randomness in the system introduced by elements $C_{ij}$, with $|i-j|\approx N$.
We observe that as we increase the value of k, the largest eigenvalue in the spectrum gradually reduces and  converges towards the bound imposed by the RMT model as shown in Fig. \ref{del}.
We then do theoretical unfolding on the new data and observe similar agreement on KS test.

For the number variance calculation, we first unfold the spectrum and calculate number variance both within bounds and over the entire spectrum. The former gives a good agreement with GUE while
the latter, as expected, shows deviation, pointing towards the presence of large eigenvalues which are due to correlation coefficients between runs scored over a long time gap.

\begin{figure}[H]
        \centering
        \includegraphics[width=0.7\textwidth]{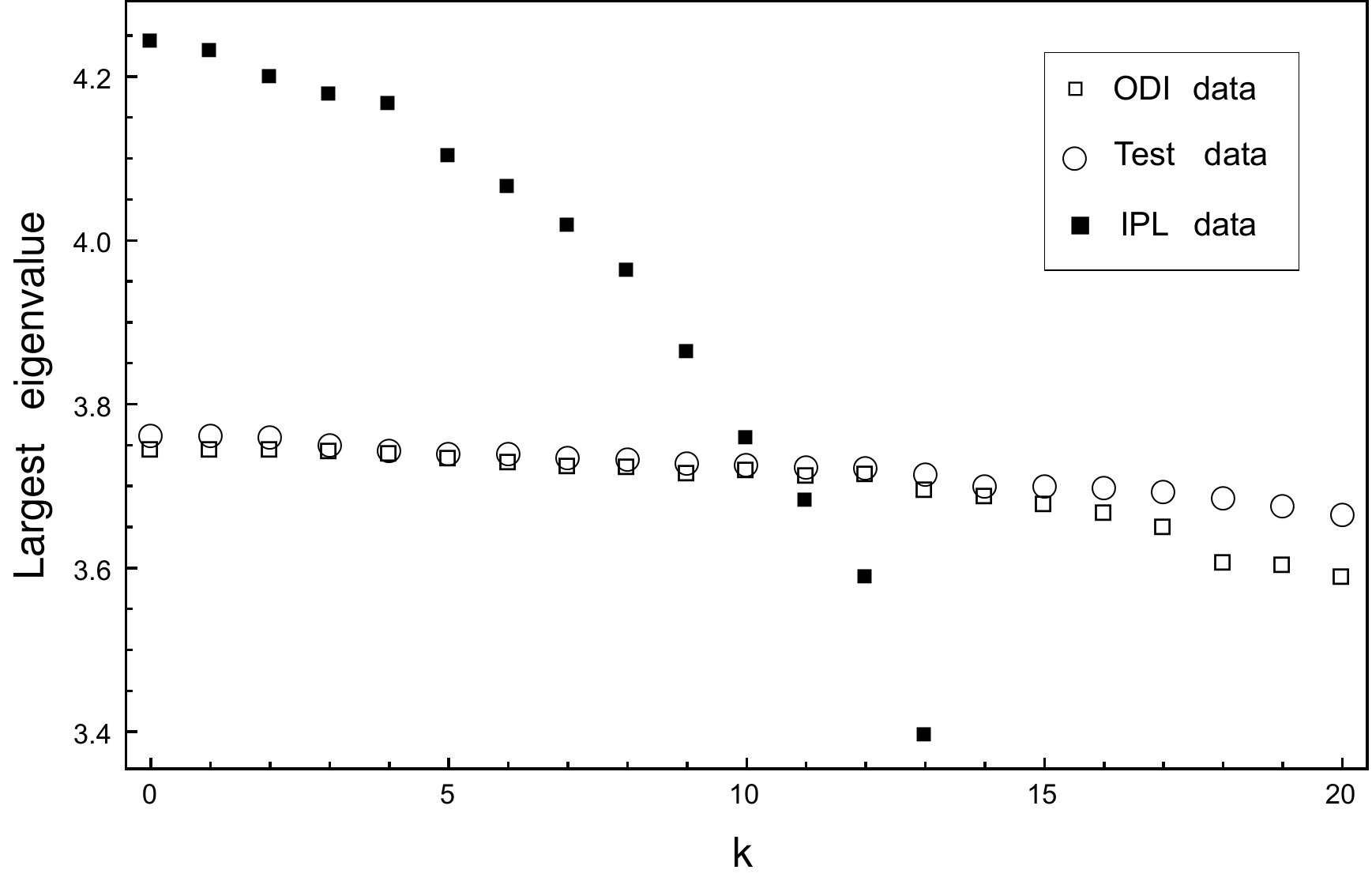}
         \caption{Largest eigenvalue in the averaged spectrum vs. $k$ for the Test, ODI and IPL data}\label{maxeigs}
\end{figure}

Finally, theoretical unfolding is performed over the spectra using Eqs.(\ref{unf}) and (\ref{den(th)}). The MP distribution parameters for the Test data ($k=5$) are given in Fig. \ref{sd}. For the ODI data ($k=15$), we have $a=2.475$, $b=3.15$, $X_{-}=0.328806$ and $X_{+}=1.87754$ as the optimal parameters for Eq. \ref{den(th)}.

Lastly, we mix levels obtained from the time series of all teams and observe a Poisson distribution (Fig. \ref{sd}).

%%%%%%%%%%%%%%%%%%%%%%%%%%%%%%%%%%%%%%%%%%%%%%%%%%%%%%%%%%%%%%%%%%%%%%%%%%%%%%%%%%%%%%%%%%%%%%%%%%%%%%%%%%%%

\section{Conclusion}
\label{sec:conclusion}
From the statistical analysis of test, ODI and IPL data, we conclude that the eigenvalues of cross-correlation matrices display GUE universality. The Test and ODI data are the only sets of data we found to be large enough to give results of the nature produced in this paper. Thus even though the T20 results of the BCCI IPL matches are also considered the small $N$ effect is visible in our GUE results.

We observe Wigner surmise when we study the ensembles of different countries (in tests and ODI s)/teams (IPL) separately. However, upon mixing the data of all countries, we get Poisson statistics, both for
spacing and number variance. Here we may recall that while studying nuclear data statistics \cite{pandey}, eigenvalues with same spin show GOE but mixed data gives Poisson.

 To ensure that the large eigenvalue which lies outside the bounds are not due to the size of the matrices, we obtain the level density using the polynomial method for finite $N$. We observe that the
 large eigenvalues were still lying  well outside the bounds. Also while numerical unfolding over the whole spectra (and not under the MP bound), we observe that the number variance show  departure from GUE. However, by removing the long-range interaction terms from $C$, we observe a better agreement with RMT predictions, both for level density as well as spacing distribution and number variance.

 We believe that eigenvalues close to the upper bound still maintains randomness and any deviation is due to temporal effect. For example, scores getting affected due to a sudden burst of performance of an individual player over a tournament or bilateral series.
However, the larger eigenvalues are probably caused due to more stable, non random influence like the effect on cricketing performance due to the advent of new technology. However this needs a thorough investigation. We wish to come back to this in a later publication.

 \section*{Acknowledgement}
We acknowledge ESPN Cricinfo for providing us with the cricket data.

\bibliographystyle{unsrt}
\bibliography{references}

\end{document}